\documentclass{article}
\usepackage{amssymb}
\usepackage{amsmath}

\begin{document}

\title{\textbf{The gravity of light\thanks{%
Corresponding author: G. Vilasi\newline
E-mail address: sparano@unisa.it, vilasi@sa.infn.it, svilasi@unisa.it, 
\newline
Research supported in part by Agenzia Spaziale Italiana (ASI) contract
WP5000 and PRIN2008 by the Italian Ministero Istruzione Universit\`{a} e
Ricerca.{}}. }\\
}
\author{G. Sparano $^{\text{a }}$, G. Vilasi $^{\text{b }}$, \ S. Vilasi $^{%
\text{c }}$ \\
%EndAName
{\small \ }$^{\text{}}${\small \ Istituto Nazionale di Fisica Nucleare,
Sezione di Napoli, Italy.}\\
{\small \ }$^{\text{a\ }}${\small \ Dipartimento di Matematica e
Informatica, Universit\`{a} di Salerno, Italy.}\\
{\small \ }$^{\text{b }}${\small \ Dipartimento di Fisica \textit{%
E.R.Caianiello}, Universit\`{a} di Salerno, Italy. }\\
{\small \ }$^{\text{c }}${\small \ Dipartimento di Scienze Farmaceutiche,
Universit\`{a} di Salerno, Italy.}\\
}
\maketitle

\begin{abstract}
A solution of the old problem raised by Tolman, Ehrenfest, Podolsky  and Wheeler, concerning the lack of attraction of two light pencils "moving parallel",  is proposed, considering that the light can be source of nonlinear gravitational waves corresponding (in the would be quantum theory of gravity) to spin$-1$ massless particles. 
\end{abstract}
 

\maketitle

\section*{Introduction}\label{sec:1}

In this paper we will study the repulsive behavior of gravitational interaction, in particle physics, 
associated to specific properties of some exact solutions of Einstein Equations. 
The interest in repulsive gravity, or {\it antigravity} as it was usually called, goes back to the fifty's \cite{MG58, Mo58, NG91}. 
The general point of view was that since gravitational interaction is mediated by a spin-2 particle, 
it can only be attractive and thus, to obtain a repulsive behavior, some other ingredient is required. 
The idea was then to explore the possibility of repulsive matter-antimatter gravity, but within the old 
quantum field theories there was no room for such a possibility. The main arguments, reviewed in \cite{NG91},  were of various kinds including violation of energy conservation and disagreement with experiments  of the E\"otv\"os type due to the effects of antigravity on the vacuum polarization diagrams of atoms. 
More recently however, within the context of modern quantum field theories, it was proven 
that those arguments were no longer sufficient to exclude repulsive effects and the interest in 
antigravity increased again. For example in \cite{FR92} it was shown that in supergravity and string theory,  due to dimensional reduction, the effective 4-dimensional theory of gravity may show repulsive  aspects because of the appearance of spin-1 graviphotons.

Our point of view is the following.
In the usual treatment of gravitational waves only Fourier expandable solutions of d'Alembert equation are considered; then it is possible to choose a special gauge (TT-gauge)  which kills the spin-0 and spin-1 components. However there exist (see section $2$ and $3$) physically meaningful solutions \cite{Pe59, St96, SKMHH03, CVV02}  of Einstein equations which are not Fourier expandable and nevertheless whose associated energy is finite.  For some of these solutions the standard analysis shows that spin-1 components cannot be killed \cite{CV04, CVV04}; this implies that repulsive aspects of gravity are possible within pure General Relativity, i.e. without involving spurious modifications.  
In previous works it was shown that light is among possible sources of such spin-1 waves \cite{CVV02, CV04, CVV04, CPV07, Vi07}. 

Photon-photon scattering can occur through the
creation and annihilation of virtual electron-positron pairs and may
even lead to collective photon phenomena. Photons also interact gravitationally
but the gravitational scattering of light by light has been much less
studied. Purely general relativistic treatments of electromagnetic
wave interactions have been made resulting in exact solutions
\cite{FPV88,FI89}, but these calculations are different from pure
scattering processes and do not address the interaction at single
photon level. It is not clear to what extent, calculations of the
gravitational cross-section using QFT methods are consistent with
classical GR. First studies go back to Tolman, Ehrenfest and Podolsky
\cite{TEP31} and, later, to Wheeler \cite{Wh55} who analysed the
gravitational field of light beams and the corresponding geodesics
in the linear approximation of Einstein equations. They discovered
that null rays behave differently according to whether they propagate
parallel or antiparallel to a steady, long, straight beam of light, but they didn't provide a physical explanation of this fact. Later,
Barker, Bathia and Gupta \cite{BBG67}, following a previous analysis of Barker, Gupta and Haracz \cite{BGH66}, analyzed in QED the photon-photon
interaction through the creation and annihilation of a virtual graviton
in the center-mass system; they found the interaction have eight times
the {}``Newtonian'' value plus a polarization dependent repulsive
contact interaction and also obtained the gravitational cross sections
for various photon polarization states. Results of Tolman, Ehrenfest, Podolsky, Wheeler were
clarified in part by Faraoni and Dumse \cite{FD99}, in the setting
of classical pure General Relativity, using an approach based on a
generalization to null rays of the gravitoelectromagnetic Lorentz
force of linearized gravity. They also extended the analysis \ to
the realm of exact \textit{pp}-wave solutions of the Einstein equations.
After Barker, Bathia and Gupta, photon-photon scattering due to self-induced
gravitational perturbations on a Minkowski background has been analyzed
by Brodin, Eriksson and Marklund \cite{BEM06} solving the Einstein-Maxwell
system perturbatively to third order in the field amplitudes and confirming
the dependence of differential gravitational cross section on the
photon polarizations.

Since the problem of the gravitational interaction of two photons
is still unsolved, it appears necessary to take into full account
the nonlinearity of Einstein's equations, just as in the case of gravitational
waves generated by strong sources \cite{Ch91,Th92}. 
Section 1 contains a short account of Einstein field equations in the linear approximation,
its gravitoelectromagnetic formulation and corresponding application to the Tolman, Erhenfest, Podolsky problem. Section 2 summarizes some geometric and physical properties (energy-momentum tensor and spin) of a family of exact solutions of Einstein equations, representing gravitational waves generated by a light beam or, more generally by massless particles. Section 3 is devoted to the motion of a massless spin-1 particle in the strong field regime. 
\section{Weak Gravitational Fields}\label{sec:2}

\subsection{The harmonic gauge} 

A gravitational field $g=g_{\mu\nu}\left(x\right)dx^{\mu}dx^{\nu}$
is said to be \textit{locally weak} if there exists a (harmonic) coordinates
system and a region $M^{\prime}\subset M$ of space-time $M$ in which
the following conditions hold: \begin{equation}
g_{\mu\nu}=\eta_{\mu\nu}+h_{\mu\nu}, \qquad\left\vert h_{\mu\nu}\right\vert <<1, \qquad\left\vert h_{\mu\nu,\alpha}\right\vert <<1.\label{debole}\end{equation}
As it is known, in the weak field approximations in a harmonic coordinates
system the Einstein field equations read \begin{equation}
\square h_{\mu\nu}=0.\label{waves}\end{equation}
The choice of the harmonic gauge plays a key role in deriving equation
(\ref{waves}); no other special assumption either on the form or
on the analytic properties of the perturbation $h$ has been done.
For globally \textit{square integrable} solutions of the wave-equation
(\ref{waves}) (that is, solutions which are square integrable
on the whole of  $M$), with a suitable gauge transformation preserving the harmonicity
of the coordinate system and the "weak character"
of the field, one can always kill the {}``spin-0'' and \textquotedblright{}spin$-1$\textquotedblright{}
components of the gravitational waves. However, in the following we
will meet some interesting solutions which do not belong to this class.

\subsection{Gravitoelectromagnetism}

A slightly different point of view, which is useful in clarifying
the nature of spin of gravitational waves is provided by the \textit{gravitoelectromagnetism},
henceforth GEM (see, for example, \cite{Ma08}). In this scheme one
tries to exploit as much as possible the similarities between the
Maxwell and the linearized Einstein equations. To make this analogy
evident it is enough to write a weak gravitational field fulfilling
conditions (\ref{debole}) in the GEM form\footnote{In this section the speed of light $c$ will be explicitly written.}.
\begin{equation}
ds^{2}=c^{2}(1+2\frac{\Phi^{\left(g\right)}}{c^{2}})dt^{2}+\frac{4}{c}({\bf A}^{\left(g\right)}\cdot d{\bf x})dt-(1-2\frac{\Phi^{\left(g\right)}}{c^{2}})\delta_{ij}dx^{i}dx^{j},\label{GEM metric}\end{equation}
with \[
h_{00}=\frac{2\Phi^{\left(g\right)}}{c^{2}}, \qquad h_{0i}=-\frac{2A_{i}^{\left(g\right)}}{c^{2}}.\]
Hereafter the spatial part of four-vectors will be denoted in bold
and the standard symbols of three-dimensional vector calculus will
be adopted. In terms of $\Phi^{\left(g\right)}$ and $\mathbf{A}^{\left(g\right)}$
the harmonic gauge condition reads \begin{equation}
\frac{1}{c}\frac{\partial\Phi^{\left(g\right)}}{\partial t}+\frac{1}{2}\nabla\cdot\mathbf{A}^{\left(g\right)}=0,\label{peppeo0}\end{equation}
and, once the gravitoelectric and gravitomagnetic fields are defined
in terms of GEM potentials, as
 \begin{equation}
\mathbf{E^{\left(g\right)\text{}}}=-\nabla\Phi^{\left(g\right)}-\frac{1}{2c}\frac{\partial\mathbf{A}^{\left(g\right)}}{\partial t}, \qquad\mathbf{B^{\left(g\right)\text{}}}=\nabla\wedge\mathbf{A}^{\left(g\right)},\label{peppeo1}
\end{equation}
one finds that the linearized Einstein equations resemble Maxwell
equations. Consequently, being the dynamics fully encoded in Maxwell-like
equations, the GEM formalism describes the physical effects of the
vector part of the gravitational field. The situations which are usually
described in this formalism are, typically, static: in fact, when
this assumption is dropped, GEM gravitational waves are also possible.

Then, the gravitoelectric and the gravitomagnetic components of the
metric are given by

\[
E_{\mu}^{(g)}=F_{\mu0}^{\left(g\right)}; \qquad B^{\left(g\right)\mu}=-\varepsilon^{\mu0\alpha\beta}F_{\alpha\beta}^{\left(g\right)}/2\quad,
\]
where
\begin{eqnarray*}
F_{\mu\nu}^{\left(g\right)} & = & \partial_{\mu}A_{\nu}^{(g)}-\partial_{\nu}A_{\mu}^{(g)}\\
A_{\mu}^{(g)} & = & -h_{0\mu}/2=(-\Phi^{(g)},\mathbf{A}^{(g)}).
\end{eqnarray*}

\begin{itemize}
\item The first order geodesic motion for a \textit{massive particle} moving  with velocity $v^\mu=(1, \b{v}) , \bf {|v|}<<1$, in
a light beam gravitational field characterized by  gravitoelectric $\mathbf{E}^{\left(g\right)}$ and gravitomagnetic  $\mathbf{B}^{\left(g\right)}$ fields,  is determined (at first order in $\bf{ |v|}$)  by the \textit{force}:
\[
\mathbf{f}^{\left(g\right)}=-\mathbf{E}^{\left(g\right)}-2\mathbf{v}\wedge\mathbf{B}^{\left(g\right)}.\]

\item The first order geodesic motion for a \textit{massless particle} moving  with velocity $v^\mu=(1,\bf{v}) $,  $|{\bf v}|=1$,
in the light beam gravitational field, parallel(anti) to $z$-axis
($v_{j}=\pm\delta_{j3}$) is slightly different 
\[
\mathbf{f}^{\left(g\right)}=-2\left(\mathbf{E}^{\left(g\right)}+\mathbf{v}\wedge\mathbf{B}^{\left(g\right)}\right).\]

The factor $2$ in front of the gravitoelectric field can be understood  as resulting from two contributions, one by the light beam, which is the source of gravity, and the other by the test photon \cite{FD99}.
It turns out  \cite{TEP31, FD99, Ze03} that for a massless particle moving parallel (antiparallel) to the light beam $\mathbf{f}^{\left(g\right)} = 0$ ($\mathbf{f}^{\left(g\right)}\neq 0$).

\end{itemize}

\section{Strong Gravitational Fields}\label{sec:3}

\subsection{Geometrical properties}

In previous papers (\cite{SVV01,SVV02a,SVV02b,CVV02,CVV04,CV04})
a family of exact solutions $g$ of Einstein field equations, representing
the gravitational wave generated by a beam of light, has been explicitly
written \begin{equation}
g=2f(dx^{2}+dy^{2})+\mu\left[(w\left(x,y\right)-2q)dp^{2}+2dpdq\right],\ \label{gm}\end{equation}
where $\mu=A\Phi+B$ with $A,B\in\mathfrak{\mathbb{R}}$, $\Phi\left(x,y\right)$
is a non constant harmonic function, $f=\left(\nabla\Phi\right)^{2}\sqrt{\left\vert \mu\right\vert }/\mu$,
and $w\left(x,y\right)$ is solution of the\textit{\ Euler-Darboux-Poisson
equation:} \[
\Delta w+\left(\partial_{x}\ln\left\vert \mu\right\vert \right)\partial_{x}w+\left(\partial_{y}\ln\left\vert \mu\right\vert \right)\partial_{y}w=\rho,\]
 $\Delta$ representing the Laplace operator in the $\left(x,y\right)-$plane
and $T_{\mu\nu}=\rho\delta_{\mu3}\delta_{\nu3}$ the energy-momentum
tensor.

It is invariant for the non Abelian Lie agebra $\mathcal{G}_{2}$
of Killing fields, generated by\[
X=\frac{\partial}{\partial p}, \qquad Y=\exp\left(p\right)\frac{\partial}{\partial q},\]
with $[X,Y]=Y$, \quad $g\left(Y,Y\right)=0$\textit{ }\textit{\emph{and
whose orthogonal distribution is integrable.}}

In the particular case $s=1$, $f=1/2$ and $\mu=1$, the above family
is locally diffeomorphic  \cite{CV04} to a subclass of 
Peres solutions \cite{Pe59,SKMHH03} and, by using the transformation\[
p=\ln\left|u\right| \qquad q=uv,\]
can be written in the form\begin{equation}
g=dx^{2}+dy^{2}+2dudv+\frac{w}{u^{2}}du^{2}\label{spin1},\end{equation}
with $\Delta w(x,y)=\rho$, and has the Lorentz invariant \textit{Kerr-Schild} form:
\[
g_{\mu\nu}=\eta_{\mu\nu}+Vk_{\mu}k_{\nu}, \qquad k_{\mu}k^{\mu}=0.
\]

\subsection{Physical Properties}

\subsubsection{Wave Character}

The wave character and the polarization of these gravitational fields
can be analyzed in many ways. For example, we could use the Zel'manov
criterion \cite{Za73} to show that these are gravitational waves
and the Landau-Lifshitz pseudo-tensor to find their propagation direction
\cite{CVV02,CVV04}. However, the algebraic Pirani criterion is easier
to handle since it determines both the wave character of the solutions
and the propagation direction at once. Moreover, it has been
shown that, in the vacuum case, the two methods agree \cite{CVV04}.
To use this criterion, the Weyl scalars must be evaluated according
to  the Newmann-Penrose formulation \cite{Pe69} of Petrov classification \cite{Pen60}.

Thus, one has to choose a
\textit{tetrad} basis with two real null vector fields and two real
spacelike (or two complex null) vector fields. Then, according to
the Pirani criterion, if the metric belongs to type \textbf{N} of
the Petrov classification, it is a gravitational wave propagating
along one of the two real null vector fields. Since $\partial_{u}$
and $\partial_{v}$ are null real vector fields and $\partial_{x}$
and $\partial_{y}$ are spacelike real vector fields, the above set
of coordinates is the right one to apply for the Pirani's criterion.

Since the only nonvanishing components of the Riemann tensor, corresponding
to the metric (\ref{spin1}), are 
\[
R_{iuju}={2 \over u^3}\partial_{ij}^{2}w(x,y), \qquad i,j=x,y\]
these gravitational fields belong to Petrov type \textbf{N }\cite{Za73}.
Then, according to the Pirani's criterion, the metric (\ref{spin1})
does indeed represent a gravitational wave propagating along the null
vector field $\partial_{u}$.

It is well known that linearized gravitational waves can be characterized
entirely in terms of the linearized and gauge invariant Weyl scalars.
The non vanishing Weyl scalar of a typical spin$-2$ gravitational
wave is $\Psi_{4}$. Metrics (\ref{spin1}) also have as non vanishing
Weyl scalar $\Psi_{4}$.

\subsubsection{Spin}

Besides being an exact solution of Einstein equations,  the metric (\ref{spin1}) 
is, for $w/u^2<<1$, also a solution of linearized Einstein equations, thus 
representing a perturbation of Minkowski metric 
$\eta$ = $dx^{2}+dy^{2}+2dudv=dx^{2}+dy^{2}+dz^{2}-dt^{2}$
(with $u=(z-t)/\sqrt2 \quad v=(z+t)/\sqrt2$) with the perturbation
\[
h:=h_{00}=h_{33}=-h_{03}=-h_{30}=\frac{w}{\left(z-t\right)^{2}}
\] 
generated by a light beam or by a photon wave packet moving along the $z$-axis.

A transparent method to determine the spin of a gravitational wave
is to look at its physical degrees of freedom, \textit{i.e.} the components
which contribute to the energy \cite{Di75}. One should use the Landau-Lifshitz
(pseudo)-tensor $t_{\nu}^{\mu}$ which, in the asymptotically flat
case, agrees with the Bondi flux at infinity \cite{CVV04}.

It is worth to remark that the canonical and the Landau-Lifchitz energy-momentum
pseudo-tensors are true tensors for Lorentz transformations. Thus, any
Lorentz transformation will preserve the form of these tensors; this
allows to perform the analysis according to the Dirac procedure. A
globally square integrable solution\textit{\ }$h_{\mu\nu}$\textit{\ }of
the wave equation is a function of\textit{\ }$r=k_{\mu}x^{\mu}$
with $k_{\mu}k^{\mu}=0.$ With the choice $k_{\mu}=(1,0,0,-1)$, we
get for the energy density \textit{\ }$t_{0}^{0}$ and the energy
momentum \textit{\ }$t_{0}^{3}$ the following result:
\[
16\pi t_{0}^{0}=\frac{1}{4}\left(u_{11}-u_{22}\right)^{2}+u_{12}^{2}, \qquad t_{0}^{0}=t_{0}^{3}
\]
where $u_{\mu\nu}\equiv dh_{\mu\nu}/d r$. Thus, the physical components
which contribute to the energy density are $h_{11}-h_{22}$ and $h_{12}$.
Following the analysis of \cite{Di75}, we see that they are eigenvectors
of the infinitesimal rotation generator $\mathcal{R}$, in the plane
$x-y,$ belonging to the eigenvalues $\pm 2i.$ The components of $h_{\mu\nu}$
which contribute to the energy thus correspond to spin$-2.$

In the case of the prototype of spin$-1$ gravitational waves (\ref{spin1}),
both Landau-Lifchitz energy-momentum pseudo-tensor and Bel-Robinson tensor \cite{Be58, Be59,Ro59}
single out the same wave components and we have: \[
\tau_{0}^{0}\sim c_{1}(h_{0x,x})^{2}+c_{2}(h_{0y,x})^{2}, \quad t_{0}^{0}=t_{0}^{3}\]
where $c_{1}$ e $c_{2}$ constants, so that the physical components
of the metric are $h_{0x}$ and $h_{0y}$. Following the previous
analysis one can see that these two components are eigenvectors of
$i\mathcal{R}$ belonging to the eigenvalues $\pm1.$ In other words,\textbf{\ }metrics
(\ref{spin1}), which are not pure gauge since the Riemann tensor
is not vanishing, represent spin$-1$ gravitational waves propagating
along the $z-$axis at light velocity.

Summarizing: \textit{globally square integrable spin}$-1$\textit{
gravitational waves propagating on a flat background are always pure
gauge. Spin}$-1$\textit{\ gravitational waves which are not globally
square integrable are not pure gauge}. It is always possible to write  metric (\ref{spin1}) 
in an apparently transverse gauge  \cite{St96}; however since these coordinates are
no more harmonic this transformation is not compatible with the linearization procedure. 

What truly distinguishes spin$-1$ from spin$-2$ gravitational waves
is the fact that in the spin$-1$ case the Weyl scalar has a non trivial
dependence on the transverse coordinates $(x,y)$ due to the presence
of the harmonic function. This could led to observable effects on
length scales larger than the \textit{characteristic length scale}
where the harmonic function changes significantly. Indeed, the Weyl
scalar enters in the geodesic deviation equation implying a non standard
deformation of a ring of test particles breaking the invariance under
of $\pi$  rotation around the propagation direction. Eventually, one
can say that there should be distinguishable effects of spin$-1$
waves at suitably large length scales.

It is also worth to stress that the results of \cite{AS71, Fe08, Ho08} suggest
that the sources of asymptotically flat PP\_waves (which have been
interpreted as spin$-1$ gravitational waves \cite{CVV02,CVV04})
repel each other. Thus, in a field theoretical perspective (see Appendix), $pp$-gravitons"
must have spin$-1$ .

\section{Back to Tolman-Erhenfest-Podolsky problem } 

In the previous strong field regime, geodesic motion can be still described in gravitoelectomagnetic formalism and  we have

\[
\mathbf{E}^{\left(g\right)}=-\frac{1}{2}(w_{x},w_{y},\frac{w}{u})u^{-2},\]

\[
\mathbf{B}^{\left(g\right)}=\frac{1}{2}(w_{y},-w_{x},\frac{w}{u})u^{-2}.\]

Thus, "gravitational force" acting over a massless particle is given
by

\begin{equation}
\mathbf{f}^{\left(g\right)}=-[w_{x}(1-v_{z})\mathbf{i}+w_{y}(1-v_{z})\mathbf{j}+(w_{x}v_{x}+w_{y}v_{y})\mathbf{k}]/2u^{2}.\label{gf}\end{equation}

Rather than geodesic orbits, the motion of spinning particles, should be described by  Papapetrou equations 
\[
{D \over D\tau}(mv^\alpha+v_\sigma{DS^{\alpha\sigma} \over D\tau})+{1 \over 2}R^\alpha_{\sigma\mu\nu}v^\sigma S^{\mu\nu}=0,
\]
where $S^{\mu\nu}$ is the {\it angular momentum tensor} of the spinning particle and $$S^\alpha={1 \over 2} \epsilon^{\alpha\beta\rho\sigma}v_\beta S_{\rho\sigma}$$ defines the {\it spin four-vector} of the particle. These equations have been extended to the case of massless spinning particles by Mashhoon {\cite{Ma75, BCGJ06}.
However, assuming that the spin is directed along the $z$-axis ${\bf S }= (0,0,S_z)$,  Papapetrou equations for photons coincide, in  the gravitational field represented by Eq.(\ref{spin1}), with usual geodesic equations and no additional contributions must be added to the "gravitational force" ${\bf f}^{(g)}$ given by Eq.(\ref{gf}).

The velocity ${\bf v}$ of  a photon is determined  by the null geodesics equations 
\[(h-1)-2hv_{z}+(h+1)v_{z}^{2}=0\]
which has two solutions\[
v_{z}=1, \qquad v_{z}=\frac{h-1}{h+1}=\frac{w-u^{2}}{w+u^{2}}\].

If  the photon propagates parallel to the light beam, $v=\left(0,0,1\right)$,
then\[
\mathbf{f}^{\left(g\right)}=0\]
and there is not attraction or repulsion.

If the photon propagates antiparallel to the light beam, $v=\left(h-1\right)/\left(h+1\right),$
then \[
\mathbf{f}^{\left(g\right)}=-\nabla w/2\left(w+u^{2}\right)\]
and the force turns out to be attractive.

\section*{Conclusions}

Thus, the lack of attraction found by Tolman, Ehrenfest, Podolsky (later also analysed by Wheeler, Faraoni and Dumse) comes out also from the analysis of the geodesical motion of a massless spin-1 test particle in the strong gravitational field of the light, neglecting however the gravitational field generated by that particle. An exhaustive answer could derive only determining the gravitational field generated by two photons, each one generating spin$-1$ gravitational waves. However, since helicity seems to play  for photons the same role that charge plays for charged particles, two photons with the same helicity  should repel one another. This repulsion turns out to be very weak and  cannot be certainly observed in laboratory but it could play a relevant role at cosmic scale and could give not trivial contributions to the dark energy.

Therefore, together with gravitons (spin-2), one may postulate the existence of 
graviphotons (spin-1) and  graviscalar (spin-0). Through coupling to fermions,
they might give forces depending on the barion number. These fields might give \cite{STM87} two
(or more) Yukawa type terms of different signs, corresponding to repulsive
graviphoton exchange and attractive graviscalar exchange (range $>> 200m$).

\section*{Acknowledgments}
One of authors (GV) wishes to thank A.P Balachandran and V. Ferrari for interesting remarks. This work is partially supported by Agenzia Spaziale Italiana (ASI), and by the Italian Ministero Istruzione Universit\`{a} e Ricerca (MIUR) through the PRIN 2008 grant.

\end{document}